\documentclass[conference]{IEEEtran}
\IEEEoverridecommandlockouts
\usepackage{cite}
\usepackage{amsmath,amssymb,amsfonts}
\usepackage{algorithmic}
\usepackage{graphicx}
\usepackage{textcomp}
\usepackage[ruled,vlined,linesnumbered,resetcount]{algorithm2e}
\usepackage{xcolor}
\def\BibTeX{{\rm B\kern-.05em{\sc i\kern-.025em b}\kern-.08em
		T\kern-.1667em\lower.7ex\hbox{E}\kern-.125emX}}
\begin{document}
	
	\title{Minimizing Energy Consumption for End-to-End Slicing in 5G Wireless Networks and Beyond\\
			\thanks{This work is supported by the Academy of Finland: (a) ee-IoT n.319009, (b) EnergyNet n.321265/n.328869, and (c) FIREMAN n.326270/CHISTERA-17-BDSI-003; and by JAES Foundation via STREAM project.}
		}
	
\author{Shiva Kazemi Taskou$ ^* $,  Mehdi Rasti$ ^{*,\dagger} $, and Pedro H. J. Nardelli$ ^\dagger $\\
$ ^* $Department of Computer Engineering, Amirkabir University of Technology, 
Tehran, Iran\\
$ ^\dagger $School of Energy Systems, {Lappeenranta-Lahti University of Technology}, Lappeenranta, Finland \\
Email: {\{shiva.kt,rasti\}@aut.ac.ir}, pedro.nardelli@lut.fi}
	
	\maketitle
	
	\begin{abstract}
		End-to-End (E2E) network slicing enables wireless networks to provide diverse services on a common infrastructure. Each E2E slice, including resources of radio access network (RAN) and core network, is rented to mobile virtual network operators (MVNOs) to provide a specific service to end-users. RAN slicing, which is realized through wireless network virtualization, involves sharing the frequency spectrum and base station antennas in RAN. Similarly, in core slicing, which is achieved by network function virtualization, data center resources such as commodity servers and physical links are shared between users of different MVNOs. In this paper, we study E2E slicing with the aim of minimizing the total energy consumption. The stated optimization problem is non-convex that is solved by a sub-optimal algorithm proposed here.  The simulation results show that our proposed joint power control, server and link allocation (JPSLA)  algorithm achieves  30\%  improvement compared to the disjoint scheme, where RAN and core are sliced separately.
	\end{abstract}
	
	
	\section{Introduction}
	Network slicing (NS) allows 5G wireless networks and beyond to offer diverse services and applications on a common infrastructure \cite{Ha-2017}. Through NS, a mobile network operator (MNO) slices its resources and infrastructure and leases them to mobile virtual network operators (MVNOs) who are responsible for providing a specific service to end-users \cite{Ha-2017}. In previous generations of wireless networks, radio access network (RAN) was a bottleneck to assure the quality of service (QoS), but with the advancement of RAN technologies in 5G and beyond, QoS is affected by the allocation of resources in both RAN and core. Hence, due to the end-to-end (E2E) concept of QoS, such as E2E delay in beyond 5G and 6G, NS should be performed in an E2E manner \cite{comst-2018}. As stated in \cite{ericson}, in wireless networks, a half of tolerable E2E delay is related to RAN and the other half is related to core network. An E2E slice is a logical network consisting of resources of RAN, backhaul link, and resources of core network rent to an MVNO to provide a specific service to end-users \cite{comst-2018}.
	
	Wireless network virtualization (WNV) and network function virtualization (NFV) are key technologies to realize RAN and core slicing, respectively.  Employing WNV in RAN, spectrum frequency and base station antennas owned by MNO are shared between different slices  \cite{Ekram}. After transmitting users' data to core networks,  the required network functions, such as packet inspection, packet routing and forwarding, etc., should be performed on users' data packets. In recent years, NFV technology has been proposed for core slicing, in which network functions are decoupled from the hardware and run as virtual network functions (VNFs) on commodity servers in data centers \cite{survey-NFV-RA}. In NFV, a sequence of  VNFs connected via virtual links forms a service function chain (SFC). In NFV, data center resources, including server capacity and physical link bandwidth, are allocated to SFCs \cite{survey-NFV-RA}.
\subsection{Related Works}	
	Some existing works studied slicing in RAN \cite{WNV-TVT-2018,co-exist-arxiv-2020,RAN-access-2020} and core network \cite{NFV-lcomm,rw-nfv-1,rw-nfv-5,nfvchain-tsc-2021}.  For instance, in  \cite{WNV-TVT-2018}  a power control problem to maximize energy efficiency subject to the users'	minimum data rate and delay constraints is addressed using fractional programming and Lyapunov optimization.  Also, the authors  in \cite{co-exist-arxiv-2020} and \cite{RAN-access-2020} studied the co-existence of enhanced mobile broadband (eMBB) and ultra-reliable low latency communication (URLLC) slices. Specifically, in \cite{co-exist-arxiv-2020} a deep reinforcement learning framework is proposed to maximize the total data rate of eMBB slices such that URLLC users' reliability is guaranteed.  Similarly, in \cite{RAN-access-2020}, a heuristic based sub-channel allocation algorithm is proposed  to slices in	such a way that the minimum data rate is satisfied for eMBB and URLLC users. 
	
	Furthermore, core slicing for assuring QoS is considered in \cite{NFV-lcomm}--\cite{nfvchain-tsc-2021}. Particularly, the authors in \cite{NFV-lcomm} proposed a heuristic algorithm for allocating the servers and links to SFCs to minimize the cost of servers and links. Besides, E2E delay minimization problem has been studied in \cite{rw-nfv-1}, where waiting delay in commodity servers queue is considered as  E2E delay. 	
	Moreover,  in \cite{rw-nfv-5}, a heuristic algorithm is proposed to minimize the number of utilized servers such that a maximum tolerable delay is guaranteed for SFCs. Additionally, authors in \cite{nfvchain-tsc-2021} proposed a heuristic algorithm for minimizing the energy consumption and cost of utilized resources in a data center.
	\subsection{Contributions and Paper Structure}
	To the best of our knowledge, none of the existing work has studied E2E slicing through which RAN and core resources are jointly allocated to the end-users. Moreover, due to high energy consumption in data centers and the environmental impacts of energy, in this paper, we study E2E slicing with the aim of minimizing total energy consumption.
	
	The remainder of this paper is structured as follows. In Section \ref{II},  the system model and notations are introduced. The optimization problem is formally stated in Section \ref{III}.  In Section \ref{IV}, we propose an algorithm to address the stated problem. Finally, the simulation results and conclusion are presented in Section \ref{V} and Section \ref{VI}, respectively.
	\section{System Model and Notations}\label{II}
	Consider a multi-cell wireless network, where an MNO rents its infrastructure and spectrum to some MVNOs\footnote{Hereafter, MVNO and slice are used interchangeably.} providing specific service to end-users. Actually, MNO slices its resources in an E2E manner,  then rents each E2E slice, including RAN and Core resources, to an MVNO. In more detail, in RAN, each user of each MVNO transmits his data to his serving base station (BS). To do so, the spectrum frequency and BSs owned by MNO are shared between users of MVNOs. The BSs transmit the received data to the core through a backhaul link. At the core, some network functions such as packet inspection, packet routing and forwarding, etc., are performed on each packet of users' data. To this end, at the core designed based on NFV, resources of data centers, including commodity servers and physical links, are shared between users of different MVNOs.
	
	In RAN, BSs and spectrum frequency are shared among users of different slices. We assume that a set of $ \mathcal{B}=\{1,2,\cdots,B\} $ BSs  serve $ G $ slices each of them provides service to $ U_g $ end-users, where the total number of users is equal to $ U=\sum\limits_{i\in \mathcal{G}} U_g $ (the set of users is denoted by $ \mathcal{U}=\{1,2,\cdots,U\} $). Besides, the whole spectrum frequency of $ W $Hz is equally divided into a set of $ \mathcal{C}=\{1,2,\cdots,C\} $ sub-channels. 
	
	If $ p_i^k $ denotes the transmit power of user $ i $ on sub-channel $ k $ and $ h_{i,m}^k $ represents the path-gain from user $ i $ to BS $ m $, the received SINR of user $ i $ on sub-channel $ k $ is given by	$ \gamma_i^k=\dfrac{p_i^k h_{i,b_i}^k}{\sum\limits_{s\notin \mathcal{U}^{b_i}}  p_s^k h_{s,b_i}^k+\sigma^2_{b_i}} $, in which $b_i  $ denotes the  BS associated with user $ i $, $ \mathcal{U}^{b_i} $ represents the users served by BS $ b_i $, and $ \sigma^2_{b_i} $ is the noise power at BS $ b_i $. 
	
	Let $ w^k= {W}/{C} $ be  bandwidth of sub-channel $ k $, the achieved data rate of user $ i $ based on Shannon's capacity formula is obtained by 
	\begin{equation}\label{data rate}
	R_i^k=w^k \log_2\left(1+\gamma_i^k\right)
	\end{equation}
	
	After the user has successfully sent his data to his serving BS, users' data are sent to the core for performing necessary network functions.   For  core network, there is a data center modeled by a directed graph as $ Graph=(\mathcal{N},\mathcal{L}) $, where $ \mathcal{N}  $ and $ \mathcal{L} $ respectively denote the set of commodity servers and physical links. For practicality, we assume servers and physical links have limited capacity denoted by $ C_n^{\mathrm{max}} $  and $ B_l^{\mathrm{max}} $, respectively. 
	
	Corresponding to the requested service, each user has an SFC represented by $ \mathcal{S}_i=\{1,2,\cdots,J_i\} $ in which  $ \mathcal{S}_i[j] $ denotes $ j $th VNF of user $ i $'s SFC.  We define binary variable $ x_{i,j}^n $ for representing the allocation of server $ n $ to $ j $th VNF of SFC $ \mathcal{S}_i $. Likewise, binary variable  $ y_{l_j^{j+1}}^i $   denotes the allocation of physical link  $ l $ to connect servers that performing $ j $th and $ j+1 $th VNFs of SFC $ \mathcal{S}_i $.
	\subsection{Calculation of End-to-End Delay and Energy Consumption}
	
	In this paper, we aim at minimizing E2E energy consumption. Similar to \cite{energy_delay-power product_2}, we define E2E energy consumption as product of delay and power. So, to calculate E2E energy consumption, we need to define E2E delay. E2E delay in cellular networks consists of delays in RAN, backhaul link, core, and transport networks \cite{latency-survey}. To obtain E2E delay, we calculate one-way delay as $ T_i=T_i^{\mathrm{RAN}}+T_i^{\mathrm{bh}}+T_i^{\mathrm{CN}}+T_i^{\mathrm{TN}} $, where $ T_i^{\mathrm{RAN}} $, $ T_i^{\mathrm{bh}} $, $ T_i^{\mathrm{CN}} $, and $ T_i^{\mathrm{TN}} $ denote delay in RAN, backhaul link, core, and transport networks \cite{latency-survey}, respectively.  It is worth noting that E2E delay is approximately obtained by $ 2\times T_i $ \cite{latency-survey}. In what follows, we explain each component of $ T_i $ in more detail. 
	
	Delay in RAN, $ T_i^{\mathrm{RAN}} $, is consisting of propagation delay, transmission delay, processing delay in users' device and BS, and queuing delay \cite{latency-survey}. In this paper, for simplicity, we consider  propagation delay,  processing delay in users' device and BS, and queuing delay as a constant value denoted by $ \tau_i $. The transmission delay is obtained by $ \dfrac{D_i}{\sum\limits_{k\in \mathcal{C}} R_i^k} $, where $ D_i $ is size of each packet in bits. Hence, delay in RAN is calculated by $ T_i^{\mathrm{RAN}}=\tau_i+\dfrac{D_i}{\sum\limits_{k\in \mathcal{C}} R_i^k} $.
	
	We assume the backhaul link capacity share among users based on the statistical multiplexing \cite{backhaul-capacity}. Using statistical multiplexing, users transmit data simultaneously over a link \cite{backhaul-capacity}. Accordingly, the delay for transmitting a packet over backhaul link is obtained by  $ T_i^{\mathrm{bh}}=\dfrac{D_i}{ C_{\mathrm{bh}}^{\mathrm{max}}} $, where $ C_{\mathrm{bh}}^{\mathrm{max}} $ denotes the maximum capacity of backhaul link. 	
	
	In core, $ T_i^{\mathrm{CN}} $ includes two components, namely processing delay and transmitting delay over physical links.  The processing delay is defined as delay	to process users' data packets on selected servers and transmitting delay is  delay for transmitting each packet between servers that perform necessary VNFs on the corresponding packet. Assume $ C_i $ is CPU cycles required by each bit of a packet, the processing delay to process a data packet of user $ i $  is obtained by $ \sum\limits_{n\in\mathcal{N}}\sum\limits_{j\in\mathcal{S}_i}x_n^{i,j}C_i D_i \left({1}/{C^{\mathrm{max}}_n}\right)$.    Note that $ {1}/{C^{\mathrm{max}}_n} $ is delay for processing one bit of each packet on server $ n $. Similar to backhaul delay, we assume the bandwidth of each link $ l $ is shared among users according to statistical multiplexing. So, delay for transmitting  a packet on physical links is calculated by $ \sum\limits_{l\in\mathcal{L}}\sum\limits_{j\in\mathcal{S}_i\backslash\{J_i\}}y^i_{l_{j}^{j+1}}D_i\left({1}/ {B_l^{\mathrm{max}}}\right)$, where $ {1}/ {B_l^{\mathrm{max}}} $ is transmission delay to transmit one bit over physical link $ l $.

	Delay on transport networks is defined as the delay for transmitting a packet from core network to the external networks such as Internet. In recent years, some architecture based on NFV and software defined networking is proposed to reduce transport delay \cite{latency-survey}. Investigation of the external data networks is out of the scope of	 this paper, so we consider a constant value as transport delay. 
	
	Based on the above discussion, the one-way delay for each data packet of user $ i $ is expressed as \eqref{DP-delay}.

	\begin{equation}\label{DP-delay}
	\begin{aligned}
	&T_i={\tau_i+\dfrac{D_i}{\sum\limits_{k\in \mathcal{C}} R_i^k}}+{\dfrac{D_i}{ C_{\mathrm{bh}}^{\mathrm{max}}}}+  {\sum\limits_{n\in\mathcal{N}}\sum\limits_{j\in\mathcal{S}_i}x_n^{i,j}C_i D_i \left({1}/{C^{\mathrm{max}}_n}\right)}\\&{+\sum\limits_{l\in\mathcal{L}}\sum\limits_{j\in\mathcal{S}_i\backslash\{J_i\}}y^i_{l_{j}^{j+1}}D_i\left({1}/ {B_l^{\mathrm{max}}}\right)}+T_i^{\mathrm{TN}}.
	\end{aligned}
	\end{equation}
	
	As aforementioned, E2E energy consumption is obtained by multiplying each component of E2E delay and its corresponding consumed power. Similar to E2E delay, E2E energy consumption consists of energy consumption in RAN, backhaul link, core, and transport networks as $ E_i=E_i^{\mathrm{RAN}}+E_i^{\mathrm{bh}}+E_i^{\mathrm{CN}}+E_i^{\mathrm{TN}} $. Considering $ p_i^k $ as transmit power of user $ i $ for transmitting data to BS and    constant energy consumption for other components denoted by $ E_i^{\mathrm{cons}} $, energy consumption in RAN is obtained by $ E_i^{\mathrm{RAN}}=\sum\limits_{k\in \mathcal{C}}  p_i^k \dfrac{D_i}{\sum\limits_{k\in \mathcal{C}}R_i^k}+E_i^{\mathrm{cons}} $.
	
	Let $ {p}_{b_i}^{\mathrm{bh}} $ denote the power consumption of BS $ b_i $ for transmitting a data packet of user $ i $ to core network through backhaul link. The energy consumption at backhaul is given by $ E_i^{\mathrm{bh}}={p}_{b_i}^{\mathrm{bh}}\dfrac{D_i}{ C_{\mathrm{bh}}^{\mathrm{max}}} $. 
	
	If $ \widetilde{p}_{n} $ represents the power consumption of server $ n $ in data center to process every CPU cycle, the energy consumption in core is obtained by $ E_{i}^{\mathrm{CN}}=\sum\limits_{n\in\mathcal{N}}\sum\limits_{j\in\mathcal{S}_i}\widetilde{p}_{n} x_n^{i,j}C_i D_i \left({1}/{C^{\mathrm{max}}_n}\right)$. Note that similar to \cite{nfvchain-tsc-2021}, it is assumed that there is no energy consumption on physical links in data centers.
	Finally, we define constant value of  $ E_i^{\mathrm{TN}} $ as energy consumption of transport networks.
	
	Accordingly, E2E energy consumption to transmit each packet of user $ i $ is obtained by $ E_i={E_i^{\mathrm{cons}}+\sum\limits_{k\in \mathcal{C}}  p_i^k \dfrac{D_i}{\sum\limits_{k\in \mathcal{C}}R_i^k}}+{{p}_{b_i}^{\mathrm{bh}}\dfrac{D_i}{ C_{\mathrm{bh}}^{\mathrm{max}}}}+{\sum\limits_{n\in\mathcal{N}}\sum\limits_{j\in\mathcal{S}_i}\widetilde{p}_{n} x_n^{i,j}C_i D_i \left({1}/{C^{\mathrm{max}}_n}\right)}+E_i^{\mathrm{TN}}$.
	It should be noted that since terms $ E_i^{\mathrm{cons}} $, $ E_i^{\mathrm{bh}} $, and $ E_i^{\mathrm{TN}} $   are not affected by decision variable i.e., $ p_i^k $, $ x_n^{i,j} $, and $ y^i_{l_j^{j+1}} $, we exclude these terms from user $ i $'s E2E energy consumption. Hence, hereafter, E2E energy consumption of user $ i $ is expressed as 
	$ E_i=\sum\limits_{k\in \mathcal{C}}  p_i^k \dfrac{D_i}{\sum\limits_{k\in \mathcal{C}}R_i^k} + \sum\limits_{n\in\mathcal{N}}\sum\limits_{j\in\mathcal{S}_i}\widetilde{p}_{n} x_n^{i,j}C_i D_i \left({1}/{C^{\mathrm{max}}_n}\right) $.
	
	\section{Problem Statement}\label{III}
	Given a sub-channel allocation, the problem of minimizing E2E energy consumption subject to users' E2E delay constraint  is formally stated as	
	\begin{equation*} 
	\begin{aligned}
	&\displaystyle \min_{\substack{\{p_i^k, x_n^{i,j},y^i_{l_j^{j+1}}\}}}
	\!\!\sum\limits_{i\in \mathcal{U}}\!\!\left(\!\sum\limits_{k\in \mathcal{C}}  p_i^k \!\!\dfrac{D_i}{\sum\limits_{k\in \mathcal{C}}R_i^k}\!\! +\!\! \sum\limits_{n\in\mathcal{N}}\sum\limits_{j\in\mathcal{S}_i}\widetilde{p}_{n} x_n^{i,j}C_i D_i\!\! \left(\!\!\dfrac{1}{C^{\mathrm{max}}_n}\!\!\right)\!\!\right)\\
	&\text{s.t.}\\
	&\mathrm{C1:} ~~\sum\limits_{k\in\mathcal{C}} p_i^k \leq p_i^{\mathrm{max}},~~~~~\forall i\in\mathcal{U},
	\end{aligned}
	\end{equation*}
	\begin{equation} \label{main-problem}
	\begin{aligned}
	&
	\mathrm{C2:} ~~\sum\limits_{i\in\mathcal{U}}\sum\limits_{k\in\mathcal{C}} R_i^k \leq {C_{\mathrm{bh}}^{\mathrm{max}}},\\
	&
	\mathrm{C3:}\sum\limits_{n\in\mathcal{N}} x_n^{i,j}=1,~~~\forall i\in\mathcal{U},~~~\forall j\in\mathcal{S}_i,\\
	&
	\mathrm{C4:}\sum\limits_{j\in\mathcal{S}_i} x_n^{i,j}\leq 1,~~~\forall i\in\mathcal{U},~~~\forall n\in\mathcal{N},\\
	&
	\mathrm{C5:}~~\sum\limits_{i\in\mathcal{U}} \sum\limits_{j\in\mathcal{S}_i}x_n^{i,j} C_i\sum\limits_{k\in \mathcal{C}}  R_i^k\leq C_{n}^{\mathrm{max}}, ~\forall n\in\mathcal{N},\\
	&
	\mathrm{C6:}~~\sum\limits_{i\in\mathcal{U}}\sum\limits_{j\in\mathcal{S}\backslash\{J_i\}}y^{i}_{l_j^{j+1}}\sum\limits_{k\in \mathcal{C}}  R_i^k\leq B_{l}^{\mathrm{max}}, ~\forall l\in\mathcal{L},\\
	&
	\mathrm{C7:}~~\sum\limits_{l\in\mathcal{L}^{\mathrm{out}}_{n}}y^i_{l_j^{j+1}}-\sum\limits_{l\in\mathcal{L}^{\mathrm{in}}_{n}}y^i_{l_j^{j+1}}=x_n^{i,j}-x_n^{i,j+1}, \\&
	~~~~~~~~~~~~~~~~ \forall i \in \mathcal{U}, \forall n\in\mathcal{N}, \forall j\in\mathcal{S}_i\backslash\{J_i\},\\
	&
	\mathrm{C8:} ~~T_i\leq T_i^{\mathrm{th}},~\forall i\in\mathcal{U},\\
	&
	\mathrm{C9:}~ p_i^k\geq 0,~~~~~~~~\forall i\in\mathcal{U},~\forall k\in\mathcal{C},\\
	&
	\mathrm{C10:}~x_n^{i,j} \in \{0,1\},~~~\forall i\in\mathcal{U},
	~~~\forall j\in\mathcal{S}_i\backslash\{J_i\},~\forall n \in\mathcal{N},\\
	&
	\mathrm{C11:}~y^{i}_{l_j^{j+1}} \in \{0,1\},~~~\forall l\in\mathcal{L},~\forall i\in\mathcal{U},~~~\forall j\in\mathcal{S}_i\backslash\{J_i\}.	
	\end{aligned}
	\end{equation}

	In problem \eqref{main-problem}, constraint $ \mathrm{C1} $ implies that the total transmit power of user $ i $ is limited to the maximum power budget  $ p_i^{\mathrm{max}} $. $ \mathrm{C2} $ indicates that the total rate of users does not exceed the maximum capacity of backhaul link. $ \mathrm{C3} $ imposes that only one server should be allocated to each VNF $ j\in\mathcal{S}_i $ for each user $ i $. $ \mathrm{C4} $ represents that every VNF for  each user $ i $ should be mapped to a different server. Constraints $ \mathrm{C5} $ and $ \mathrm{C6} $ demonstrate the capacity limitation of commodity servers and bandwidth limitation of physical links in data center, respectively. If $ \mathcal{L}_n^{\mathrm{out}} $ and $ \mathcal{L}_n^{\mathrm{in}} $ denote the outgoing links from server $ n $ and incoming links to server $ n $, respectively, constraint $ \mathrm{C7} $ enforces flow conservation, i.e., the sum of all incoming and outgoing traffic in  servers that do not host VNFs should be zero.  $ \mathrm{C8} $ implies that E2E delay of user $ i $ should not be larger than maximum tolerable delay denoted by $ T_i^{\mathrm{th}} $.  $ \mathrm{C9} $ represents the non-negativity of transmit power on each sub-channel. Finally,  $ \mathrm{C10} $ and  $ \mathrm{C11} $ indicate the binary nature of server and physical link allocation, respectively. 
	
	\section{Our Proposed JPSLA Algorithm } \label{IV}
	Due to (i) non-concavity of data rate function \eqref{data rate}, (ii) existence of continuous and integer variables, and (iii) multiplying of transmit power and transmission delay (i.e., $ \sum\limits_{k\in \mathcal{C}}  p_i^k \dfrac{D_i}{\sum\limits_{k\in \mathcal{C}}R_i^k} $) in objective function, the optimal solution of problem \eqref{main-problem} cannot be obtained in polynomial time. So, in what follows, we propose a sub-optimal joint power control, server and link allocation (JPSLA) to solve problem \eqref{main-problem}. 
	
	To tackle with existence of continuous and integer variables, we decompose problem \eqref{main-problem} into two sub-problems, namely power control problem \eqref{power control} in RAN and server and link allocation problem \eqref{server-link allocation} in core as 
	
	\begin{equation}\label{power control}
	\begin{aligned}
	&\displaystyle \min_{\substack{\{p_i^k\}}}
	&& \sum\limits_{i\in \mathcal{U}}\sum\limits_{k\in \mathcal{C}}  p_i^k \dfrac{D_i}{\sum\limits_{k\in \mathcal{C}}R_i^k} \\
	&\text{s.t.}
	&&\mathrm{C1}, \mathrm{C2}, \mathrm{C5}, \mathrm{C6}, \mathrm{C8},\mathrm{C9},	
	\end{aligned}
	\end{equation}
	and
	\begin{equation}\label{server-link allocation}
	\begin{aligned}
	&\displaystyle \min_{\substack{\{ x_n^{i,j},y^i_{l_j^{j+1}}\}}}
	&& \sum\limits_{i\in \mathcal{U}} \sum\limits_{n\in\mathcal{N}}\sum\limits_{j\in\mathcal{S}_i}\widetilde{p}_{n} x_n^{i,j}C_i D_i \left({1}/{C^{\mathrm{max}}_n}\right)\\
	&\text{s.t.}
	&&\mathrm{C3}, \mathrm{C4}, \mathrm{C5}, \mathrm{C6}, \mathrm{C7}, \mathrm{C8}, \mathrm{C10}, \mathrm{C11}. 	
	\end{aligned}
	\end{equation}
	\subsection{Solving Power Control Problem \eqref{power control}}
	Power control problem \eqref{power control} is a non-convex problem due to (i) non-convexity of constraints $ \mathrm{C2} $,  $ \mathrm{C5} $, and $ \mathrm{C6} $, (ii) multiplying of transmit power and transmission delay (i.e., $ \sum\limits_{k\in \mathcal{C}}  p_i^k \dfrac{D_i}{\sum\limits_{k\in \mathcal{C}}R_i^k} $) in objective function, and (iii) non-concavity of data rate function \eqref{data rate}. 
	
	To tackle with non-convexity of constraints $ \mathrm{C2} $,  $ \mathrm{C5} $, and $ \mathrm{C6} $, similar to \cite{auxilary}, we define $ \nu_i^k $   as an auxiliary variable and substitute $ R_i^k $ in problem \eqref{power control} by $ \nu_i^k $ and add following constraint to problem \eqref{power control}
	\begin{equation}\label{auxillary}
	\begin{aligned}
	\mathrm{C12:}~~\nu_i^k \leq  R_i^k.
	\end{aligned}
	\end{equation}
	
	While $ \mathrm{C2} $,  $ \mathrm{C5} $, and $ \mathrm{C6} $ become  convex, problem \eqref{power control} is still non-convex because of non-convexity of the objective function and  non-concavity of data rate in constraint $ \mathrm{C12} $.  	
	As aforementioned, the objective function  of problem \eqref{power control} is non-convex due to the existence of $ \sum\limits_{k\in \mathcal{C}}  p_i^k \dfrac{D_i}{\sum\limits_{k\in \mathcal{C}}{R}_i^k} $. Since the energy consumption derivative is positive with respect to the user's transmit power, implying that energy consumption is decreased by decreasing the transmit power, the minimum amount of energy consumption can be obtained by minimizing users' total transmit power. 
	Moreover, constraint $ \mathrm{C9} $ in problem \eqref{power control} indicates that user $ i $'s transmission delay to transmit a packet to BS is upper bounded by $ T_i^{\mathrm{th}}-(\tau_i+{T_i}^{ {\mathrm{bh}}}+T_i^{\mathrm{CN}}+T_i^{\mathrm{TN}})  $   i.e., $ \dfrac{D_i}{\sum\limits_{k\in \mathcal{C}}{R}_i^k}\leq T_i^{\mathrm{th}}-(\tau_i+{T_i}^{ {\mathrm{bh}}}+T_i^{\mathrm{CN}}+T_i^{\mathrm{TN}})$. Accordingly, user $ i $'s energy consumption for sending a packet to BS is approximated as  $ { \sum\limits_{k\in \mathcal{C}} p_i^k}\left(T_i^{\mathrm{th}}-(\tau_i+{T_i}^{ {\mathrm{bh}}}+T_i^{\mathrm{CN}}+T_i^{\mathrm{TN}}) \right)$.	
	
	Finally, to make  data rate function \eqref{data rate} a concave one,  we consider a predefined interference threshold $ I_i^{k,\mathrm{th}} $ instead of interference terms which make data rate function \eqref{data rate} non-concave. To this end, we add the following constraint to problem \eqref{power control} 
	\begin{equation}\label{interfence threshold}
	\begin{aligned}
	\mathrm{C13:}~~\sum\limits_{s\notin \mathcal{U}^{b_i}}  p_s^k h_{s,b_i}^k \leq I_i^{k,\mathrm{th}},~~~\forall i \in\mathcal{U},~~~\forall k \in\mathcal{C}.
	\end{aligned}
	\end{equation}
	By doing so, data rate of user $ i $ on sub-channel $ k $ is obtained by $ \widetilde{R}_i^k=w^k \log_2\left(1+\dfrac{p_i^k h_{i,b_i}^k}{I_i^{k,\mathrm{th}}+\sigma^2_{b_i}}\right) $. 	
	Employing the above discussed steps, problem \eqref{power control} is converted to 
	\begin{equation*} 
	\begin{aligned}
	&\displaystyle \min_{\substack{\{p_i^k\}}} ~~~~~
	\sum\limits_{i\in \mathcal{U}} { \sum\limits_{k\in \mathcal{C}} p_i^k}\left(T_i^{\mathrm{th}}-(\tau_i+{T_i}^{ {\mathrm{bh}}}+T_i^{\mathrm{CN}}+T_i^{\mathrm{TN}}) \right) \\
	&\text{s.t.}\\
	&\mathrm{C1:} ~~\sum\limits_{k\in\mathcal{C}} p_i^k \leq p_i^{\mathrm{max}},~~~~~\forall i\in\mathcal{U},\\
	&
	\mathrm{C2:} ~~\sum\limits_{i\in\mathcal{U}}\sum\limits_{k\in\mathcal{C}} \nu_i^k \leq {C_{\mathrm{bh}}^{\mathrm{max}}},\\
	&
	\mathrm{C5:}~~\sum\limits_{i\in\mathcal{U}} \sum\limits_{j\in\mathcal{S}_i}x_n^{i,j} C_i\sum\limits_{k\in \mathcal{C}}  \nu_i^k\leq C_{n}^{\mathrm{max}}, ~\forall n\in\mathcal{N},
	\end{aligned}
	\end{equation*}
	\begin{equation}\label{power control-converted}
	\begin{aligned}
	&
	\mathrm{C6:}~~\sum\limits_{i\in\mathcal{U}}\sum\limits_{j\in\mathcal{S}\backslash\{J_i\}}y^{i}_{l_j^{j+1}}\sum\limits_{k\in \mathcal{C}}  \nu_i^k\leq B_{l}^{\mathrm{max}}, ~\forall l\in\mathcal{L},\\
	&
	\mathrm{C8:} ~~\dfrac{D_i}{\sum\limits_{k\in \mathcal{C}}\nu_i^k}+\tau_i+T_i^{\mathrm{bh}}+T_i^{\mathrm{CN}}+T_i^{\mathrm{TN}}\leq T_i^{\mathrm{th}},~\forall i\in\mathcal{U},\\
	&
	\mathrm{C9},\mathrm{C12},\mathrm{C13}.	
	\end{aligned}
	\end{equation}
	Problem \eqref{power control-converted} is a convex, so it can be optimally solved by CVX \cite{cvx}.
	\subsection{Solving Sever and Link Allocation Problem \eqref{server-link allocation}}
	Problem \eqref{server-link allocation} is an integer linear programing problem. To handle integer variables $ x_n^{i,j} $ and $ y^i_{l_j^{j+1}} $, we replace $ \mathrm{C10} $  and $ \mathrm{C11} $ in problem \eqref{server-link allocation} by following constraints
	\begin{equation}\label{relaxation-server}
	\begin{aligned}
	&\mathrm{C10.1:}~~\sum\limits_{n\in\mathcal{N}}\sum\limits_{i\in\mathcal{U}}\sum\limits_{j\in\mathcal{S}_i}(x_n^{i,j}-{x_n^{i,j}}^2)\leq 0,\\
	&\mathrm{C10.2:}~~0\leq x_n^{i,j}\leq 1,~~~\forall n\in\mathcal{N},\forall i \in\mathcal{U},\forall j\in\mathcal{S}_i.
	\end{aligned}
	\end{equation}
	and
	\begin{equation}\label{relaxation-link}
	\begin{aligned}
	&\mathrm{C11.1:}~~\sum\limits_{l\in\mathcal{L}}\sum\limits_{i\in\mathcal{U}}\sum\limits_{j\in\mathcal{S}_i}(y^{i}_{l_j^{j+1}}-{y^{i}_{l_j^{j+1}}}^2)\leq 0,\\
	&\mathrm{C11.2:}~~0\leq y^{i}_{l_j^{j+1}}\leq 1,\forall l\in\mathcal{L},\forall i \in\mathcal{U},\forall j\in\mathcal{S}_i.
	\end{aligned}
	\end{equation}

	Adding constraints $ \mathrm{C10.2} $  and $ \mathrm{C11.2} $ to problem \eqref{server-link allocation} which relax binary variables $ x_n^{i,j} $ and $ y^i_{l_j^{j+1}} $ to continuous ones may make problem \eqref{server-link allocation} infeasible. So, for enforcing  $ x_n^{i,j} $ and $ y^i_{l_j^{j+1}} $ to binary variables, we add constraints  $ \mathrm{C10.1} $  and $ \mathrm{C11.1} $ to objective function as a penalty function. By doing so,  problem \eqref{server-link allocation} is transformed into 
	\begin{equation}\label{server-link allocation-converted}
	\begin{aligned}
	&\displaystyle \min_{\substack{\{ x_n^{i,j},y^i_{l_j^{j+1}}\}}}
	&& \sum\limits_{i\in \mathcal{U}}  \sum\limits_{n\in\mathcal{N}}\sum\limits_{j\in\mathcal{S}_i}\widetilde{p}_{n} x_n^{i,j}C_i D_i \left({1}/{C^{\mathrm{max}}_n}\right)\\ &	
	&&
	+\zeta_1\sum\limits_{n\in\mathcal{N}} \sum\limits_{i\in\mathcal{U}}\sum\limits_{j\in\mathcal{S}_i}(x_n^{i,j}-{x_n^{i,j}}^2)\\
	&
	&&+\zeta_2\sum\limits_{l\in\mathcal{L}} \sum\limits_{i\in\mathcal{U}}\sum\limits_{j\in\mathcal{S}_i}(y^{i}_{l_j^{j+1}}-{y^{i}_{l_j^{j+1}}}^2)\\
	&\text{s.t.}
	&&\mathrm{C3}, \mathrm{C4}, \mathrm{C5}, \mathrm{C6}, \mathrm{C7}, \mathrm{C8}, \mathrm{C10.2}, \mathrm{C11.2}, 	
	\end{aligned}
	\end{equation}
	where $ \zeta_1 $ and $ \zeta_2 $ are penalty factors for enforcing $ x_n^{i,j} $ and $ y^i_{l_j^{j+1}} $ to  $ 0 $ or $ 1 $. 
	
	The existence of $ {x_n^{i,j}}^2 $ and $ {y^{i}_{l_j^{j+1}}}^2 $ in the objective function make  problem \eqref{server-link allocation-converted} non-convex.  To deal with this difficulty, we approximate $ {x_n^{i,j}}^2 $ and $ {y^{i}_{l_j^{j+1}}}^2 $ by  first-order Taylor approximation approach \cite{MM-approximation}. 
	First-order Taylor approximation approximates a function  $ f(x) $ by a linear function as 
	\begin{equation}\label{taylor}
	\begin{aligned}
	f(x)\approx f(\overline{x})+\nabla_{{x}}f(\overline{x})(x-\overline{x}),
	\end{aligned}
	\end{equation}
	where, $ \overline{x} $ is a feasible initial point.
	
	Employing \eqref{taylor} to approximate $ {x_n^{i,j}}^2 $ and $ {y^{i}_{l_j^{j+1}}}^2 $, problem \eqref{server-link allocation-converted} is rewritten as  
	\begin{equation}\label{server-link allocation-converted-re}
	\begin{aligned}
	&\displaystyle \min_{\substack{\{ x_n^{i,j},y^i_{l_j^{j+1}}\}}}
	\sum\limits_{i\in \mathcal{U}}E_i\\
	&+\zeta_1\sum\limits_{n\in\mathcal{N}} \sum\limits_{i\in\mathcal{U}}\sum\limits_{j\in\mathcal{S}_i}\left(x_n^{i,j}+\big[2{x_n^{i,j}}{{{x}_n^{i,j}(t-1)}}-({{{x}_n^{i,j}(t-1)})^2}\big]\right)\\
	&+\zeta_2\sum\limits_{l\in\mathcal{L}} \sum\limits_{i\in\mathcal{U}}\sum\limits_{j\in\mathcal{S}_i}\!\!\left(y^{i}_{l_j^{j+1}}\!+\!\big[2{y^{i}_{l_j^{j+1}}}{{{y}^{i}_{l_j^{j+1}}(t-1)}}\!\!-\!\!({{{y}^{i}_{l_j^{j+1}}(t-1)})^2}\big] \!\right)\\
	&\text{s.t.}~~~~~~
	\mathrm{C3}, \mathrm{C4}, \mathrm{C5}, \mathrm{C6}, \mathrm{C7}, \mathrm{C8}, \mathrm{C10.2}, \mathrm{C11.2}, 	
	\end{aligned}
	\end{equation}
	Problem \eqref{server-link allocation-converted-re} is a linear problem that can be optimally solved by CVX \cite{cvx}.
	Note that $ {x}_n^{i,j}(t-1) $ and $ {y}^{i}_{l_j^{j+1}}(t-1) $ denote the optimal solution of problem \eqref{server-link allocation-converted-re} at the previous iteration $ t-1 $.
	
	\begin{algorithm}\label{RRA-CRA}
		\BlankLine	
		
		\footnotesize
		\SetKwFunction{Range}{range}
		\SetKw{KwTo}{in}\SetKwFor{For}{for}{\string:}{}%
		\SetKwIF{If}{ElseIf}{Else}{if}{:}{elif}{else:}{}%
		
		\SetAlgoNoEnd
		
		\SetAlgoNoLine%
		\SetKwInOut{Input}{Input}
		\SetKwInOut{Output}{Output}
		Initialize the maximum number of iterations $ t^{\mathrm{max}} $, $ \zeta_1, \zeta_2\gg 1 $,  iteration index $ t=1 $, and a feasible initial point  $ {x}_n^{i,j}(0) $,  and $ {y}^i_{l_j^{j+1}}(0) $.\\
		\textbf{Repeat }\\
		\Indp
		\textbf{Step 1: Solving power control problem \eqref{power control-converted}}\\
		\Indp
		Solve \eqref{power control-converted} by CVX to obtain $ p_i^k(t),~\forall i\in\mathcal{U},~\forall k\in\mathcal{C} $.\\
		\Indm
		\textbf{Step 2: Solving server and link allocation problem \eqref{server-link allocation-converted-re}  }\\
		\Indp
		Solve \eqref{server-link allocation-converted-re} by CVX to obtain $ {x}_n^{i,j}(t) $,  and $ {y}^i_{l_j^{j+1}}(t) $.\\
		\Indm
		Set  	 $ t\leftarrow t+1 $.\\
		\Indm
		\textbf{Until } convergence or $ t=t^{\mathrm{max}} $.
		\caption{ JPSLA algorithm  to solve  problem \eqref{main-problem}}	
	\end{algorithm}
	
	For obtaining a local optimum for problem \eqref{main-problem}, we propose JPSLA algorithm as an iterative algorithm summarized in Algorithm 1. At each iteration of JPSLA, convex problems \eqref{power control-converted} and \eqref{server-link allocation-converted-re} are solved by CVX until convergence is occurred. 
	
	\section{Simulation Results}\label{V}
	In this section, we evaluate the performance of our proposed JPSLA algorithm. To do so,   a multi-cell network consisting of $ B=2 $ BSs   with a coverage area of $ 500\mathrm{m}\times 500\mathrm{m} $ is considered.  We assume in this network, $ G=3 $ slices, for instance, factory automation, health care, and intelligent transport systems are provided to end-users.   
	Moreover, end-users of different slices are randomly distributed at each cell. Like  \cite{pathgain},  the path gain between  each user and BSs is modeled by $ h^k_{i,m}=\mu^k d_{i,m} ^{-\beta} $, where $ d_{i,m} $ is the distance between user $ i $ and BS $ m $, $ \mu^k $ is a random value generated by the Rayleigh distribution, and $ \beta=3 $ is the path loss exponent.
	Additionally, for the core network, it is assumed that commodity servers are connected to each other with randomly established physical links.   Randomly setting locations and path-gain of users in each snapshot,   each curve in what follows is obtained by averaging from  $ 50 $ independent snapshots.
	
	To obtain simulation figures,  bandwidth of each sub-channel  is set to  $ w^k=15~\mathrm{KHz} $. Furthermore, each user $ i $ should not transmit larger than $ 100~\mathrm{mW} $ (i.e., $p_i^{\mathrm{max}}=100~\mathrm{mW}$). Also, noise power at BS $ m $ is set to  $\sigma_{m}^2=10^{-14}~\mathrm{W} $.  Each user $ i $ has  packets with size $D_i =100~\mathrm{bits}$. Moreover, the capacity of backhaul link is limited to $C_{\mathrm{bh}}^{\mathrm{max}}=1~\mathrm{Gbps}$.  Besides, the maximum capacity of commodity servers and physical links are randomly selected from $ [10,50]\mathrm{MHz} $ and $ [100,500]\mathrm{Mbps} $, respectively. In addition, we set constant delay in RAN and transport network delay   to $\tau_i =2~\mathrm{ms}$ and  $T_i^{\mathrm{TN}}=1~\mathrm{ms}$, respectively. Finally, power consumption 	of each server $ n $ i,e., $ \widetilde{p}_n$	is randomly selected from $[1,10]~\mathrm{W}$.
	
	To show the performance of our proposed JPSLA algorithm, we compare it with the disjoint (DS) scheme, where the power control problem in RAN and server and link allocation in the core are solved disjointly. As stated in \cite{ericson}, half of the tolerable E2E delay for each user is related to RAN and the other half is related to the core. Firstly, in the DS scheme, the power control problem \eqref{power control} is solved considering half of the maximum tolerable delay in RAN. Then, given transmit power of users, server and link allocation problem \eqref{server-link allocation} is solved considering half of the maximum tolerable delay in the core.
	
	\begin{figure}
		\centerline{\includegraphics[width=0.40\textwidth,height=1.75in]{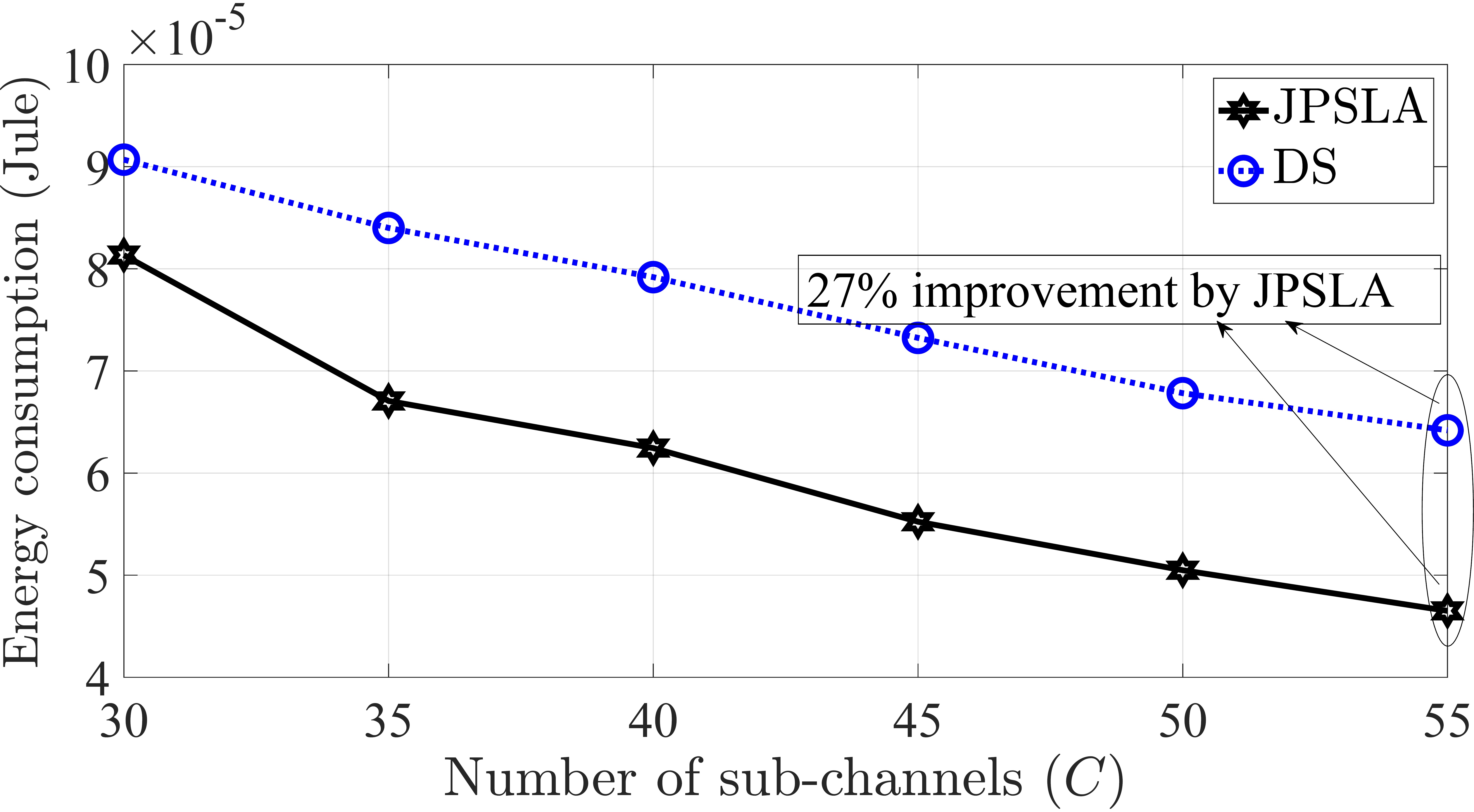}}
		\caption{\small Comparison of JPSLA   and disjoint scheme vs. sub-channels number }\label{joint_ds_subchannel}\vspace{-0.8 em}
		
	\end{figure}
	
	\begin{figure}
		\centerline{\includegraphics[width=0.40\textwidth,height=1.75in]{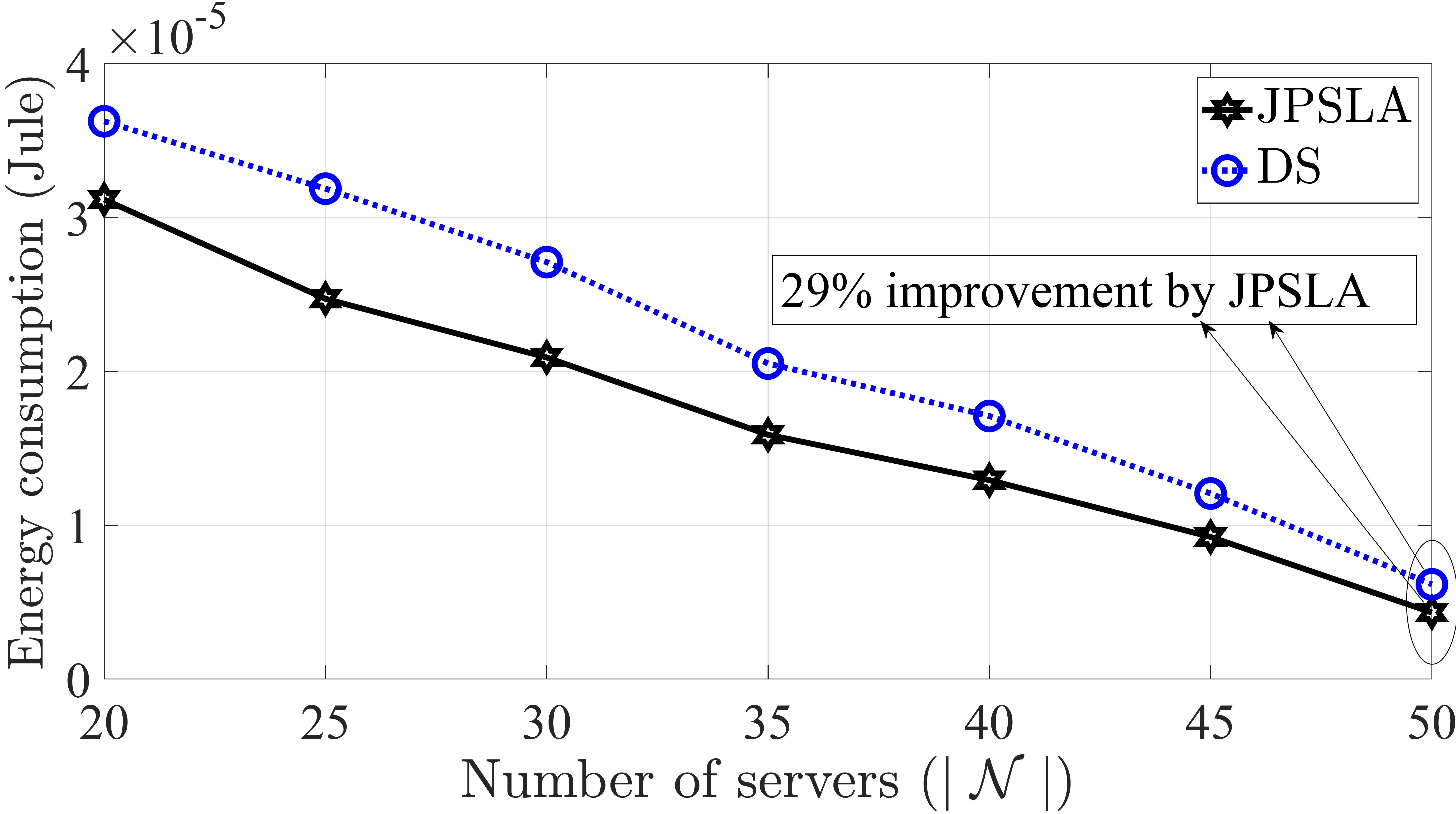} }
		\caption{\small Comparison of JPSLA   and disjoint scheme vs. servers number }\label{joint_ds_server}\vspace{-0.8 em}
		
	\end{figure}
	
	In Fig. \ref{joint_ds_subchannel}, energy consumption in JPSLA algorithm is compared with the DS scheme, where the number of sub-channels increases from $ 30 $ to $ 55 $. For generating this figure, we set number of servers to $ \mid\mathcal{N}\mid=20 $. Also, a different maximum tolerable delay is considered for different slices. Specifically, the maximum tolerable delay of users in slice $ 1 $, $ 2 $, and $ 3 $ are respectively set to $ T_i^{\mathrm{th}}=5\mathrm{ms} $, $ T_i^{\mathrm{th}}=10\mathrm{ms} $, and $ T_i^{\mathrm{th}}=15\mathrm{ms} $. Also, it is assumed that each slice provides service to $ U_g=4 $ users equally distributed in cell. It can be seen that because of joint consideration of power control, server and link allocation, as well as considering half of the tolerable delay in RAN and core in DS scheme, energy consumption in JPSLA is less than that of DS scheme.  Besides, it can be observed that due to sub-channel diversity, increasing the number of sub-channel, energy consumption is decreased.
	
	\begin{figure}
		\centerline{		\includegraphics[width=0.40\textwidth,height=1.75in] {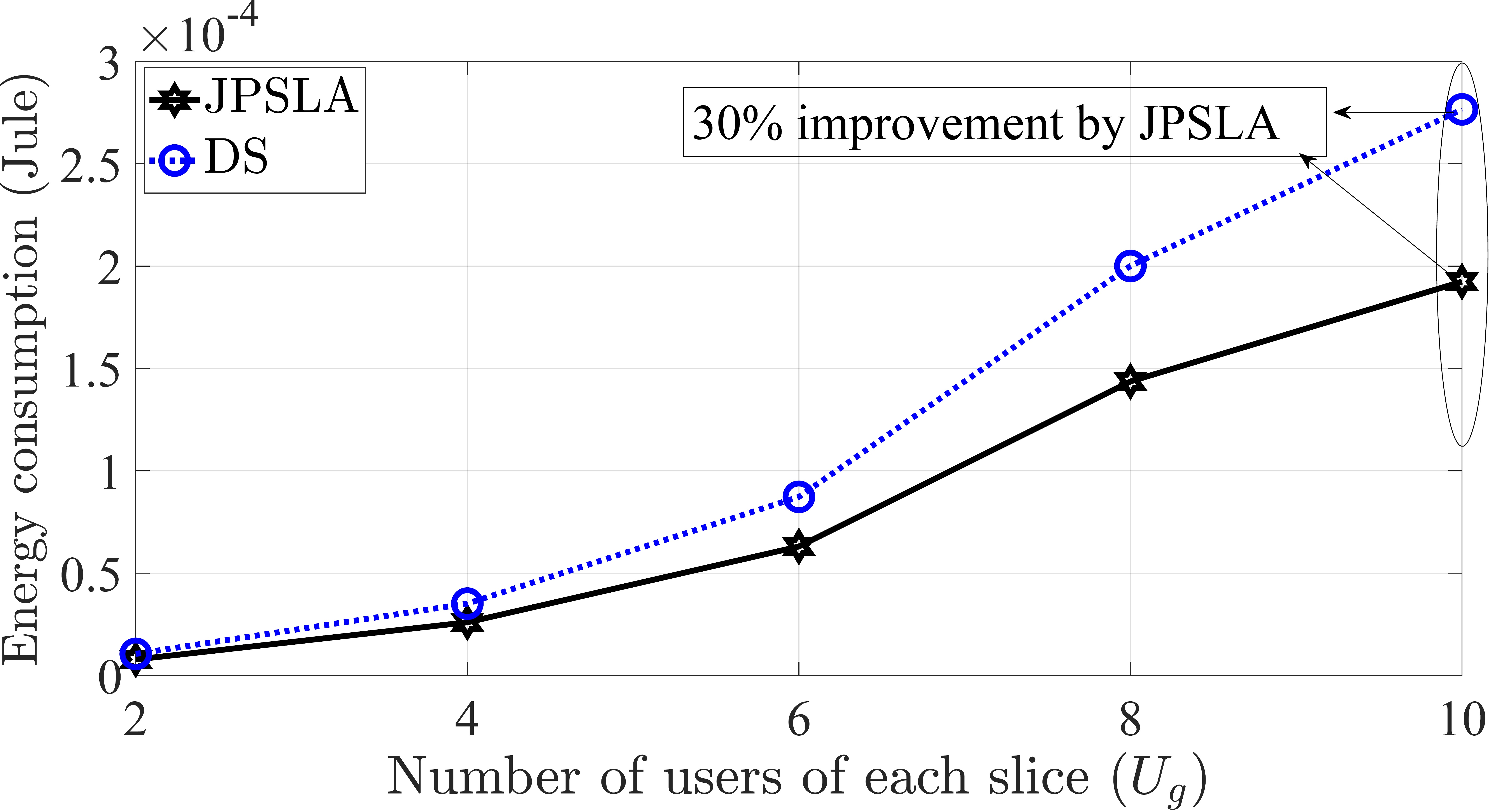}}
		\caption{\small Comparison of JPSLA  and disjoint scheme vs. users' number }\label{joint_ds_users}\vspace{-0.8 em}
		
	\end{figure}
	
	Fig. \ref{joint_ds_server} illustrates the energy consumption in JPSLA algorithm in comparison with the DS scheme. To generate this figure, we set the number of sub-channels to $ C=30 $, and the other parameters are set the same as Fig. \ref{joint_ds_subchannel}. 
	From Fig. \ref{joint_ds_server}, we can observe that with increasing the number of servers, the energy consumption is decreased. The reason is that commodity servers in the data centers are heterogeneous in terms of maximum capacity; also, since the maximum capacity of servers is set randomly, the probability of the existence of high-capacity servers is increased. Furthermore,  the energy consumption in JPSLA is lower than that of the DS scheme because of the joint allocation of decision variables. Besides,  considering half of the maximum tolerable delay in both RAN and core, the E2E delay constraint in the DS scheme is more strict than JPSLA.
	
	In Fig. \ref{joint_ds_users}, we increase the number of users served by each slice from $ U_g=2 $ to $ U_g=10 $. To generate this figure, we assume that $ C=50 $ sub-channels are equally allocated to users at each cell.  Fig. \ref{joint_ds_users} demonstrates that the energy consumption in both JPSLA and DS scheme is increased when the number of users is increased. Obviously, with increasing the number of users, since more users cause interference on each other, users should transmit with higher transmit power resulting in more energy consumption.
	
	\section{Conclusion}\label{VI}
	In this paper, we studied  E2E slicing so that RAN and core resources are jointly allocated to end-users of MVNOs. Due to the importance of energy consumption in wireless networks and its environmental impacts, we aimed at minimizing E2E energy consumption. To address the stated problem, which was non-convex, we proposed JPSLA algorithm. Via simulation results, we illustrated that JPSLA improves energy consumption by  30\%  in comparison with the disjoint scheme, where RAN and core are sliced separately.


\begin{thebibliography}{00}
		
		\bibitem{Ha-2017}
		V. N. Ha and L. B. Le, {``End-to-End Network Slicing in Virtualized OFDMA-Based Cloud Radio Access Networks,"} \textit{IEEE Access}, vol. 5, pp. 18675-18691, 2017.
		
		\bibitem{comst-2018}
		I. Afolabi, T. Taleb, K. Samdanis, A. Ksentini, and H. Flinck, {``Network Slicing and Softwarization: A Survey on Principles, Enabling Technologies, and Solutions,"}  \textit{IEEE Communications Surveys \& Tutorials}, vol. 20, no. 3, pp. 2429-2453, 2018.	
		
		\bibitem{ericson}
		White Paper, {``Cellular IoT in the 5G era,"} \textit{Ericsson}, 2020.	
		
		\bibitem{Ekram}
		K. Zhu and E. Hossain, {``Virtualization of 5G Cellular Networks as a Hierarchical Combinatorial Auction,"} \textit{IEEE Transactions on Mobile Computing}, vol. 15, no. 10, pp. 2640-2654, 2016.
		
		\bibitem{survey-NFV-RA}
		J. Gil Herrera and J. F. Botero, {``Resource Allocation in NFV: A Comprehensive Survey,"} \textit{IEEE Transactions on Network and Service Management}, vol. 13, no. 3, pp. 518-532, 2016.	
		
		\bibitem{WNV-TVT-2018}
		Q. Shi, L. Zhao, Y. Zhang, G. Zheng, F. R. Yu, and H. Chen, {``Energy-Efficiency Versus Delay Tradeoff in Wireless Networks Virtualization,"} \textit{IEEE Transactions on Vehicular Technology}, vol. 67, no. 1, pp. 837-841, 2018.
		\bibitem{co-exist-arxiv-2020}
		M. Alsenwi, N. H. Tran, M. Bennis, S. R. Pandey, A. K. Bairagi, and C. S. Hong, {``Intelligent Resource Slicing for eMBB and URLLC Coexistence in 5G and Beyond: A Deep Reinforcement Learning Based Approach,"} \textit{Available online at: }{https://arxiv.org/pdf/2003.07651.pdf}, 2020.
		\bibitem{RAN-access-2020}
		P. Korrai, E. Lagunas, S. K. Sharma, S. Chatzinotas, A. Bandi, and B. Ottersten, {``A RAN Resource Slicing Mechanism for Multiplexing of eMBB and URLLC Services in OFDMA Based 5G Wireless Networks,"} \textit{IEEE Access}, vol. 8, pp. 45674-45688, 2020.	
		\bibitem{NFV-lcomm}
		V. Eramo and F. G. Lavacca, {``Computing and Bandwidth Resource Allocation in Multi-Provider NFV Environment,"} \textit{ IEEE Communications Letters}, vol. 22, no. 10, pp. 2060-2063, 2018.
		\bibitem{rw-nfv-1}
		G. Sun, Z. Chen, H. Yu, X. Du, and M. Guizani, {``Online Parallelized Service Function Chain Orchestration in Data Center Networks,"} \textit{IEEE Access}, vol. 7, pp. 100147-100161, 2019.
		\bibitem{rw-nfv-5}
		M. Savi, M. Tornatore, and G. Verticale, {``Impact of Processing-Resource Sharing on the Placement of Chained Virtual Network Functions,"} \textit{IEEE Transactions on Cloud Computing}, 2019,   early access.
		
		\bibitem{nfvchain-tsc-2021}
		S. Kazemi Taskou, M. Rasti, and P. H. J. Nardelli, {``Energy and Cost Efficient Resource Allocation for Blockchain-Enabled NFV,"}  \textit{IEEE Transactions on Services Computing}, early access, 2021.
		
		\bibitem{energy_delay-power product_2}
		M. Chen and Y. Hao, {``Task Offloading for Mobile Edge Computing in Software Defined Ultra-Dense Network,"}  \textit{IEEE Journal on Selected Areas in Communications}, vol. 36, no. 3, pp. 587-597,  2018.
		
		\bibitem{latency-survey}
		I. Parvez, A. Rahmati, I. Guvenc, A. I. Sarwat, and H. Dai, {``A Survey on Low Latency Towards 5G: RAN, Core Network and Caching Solutions,"} \textit{IEEE Communications Surveys \& Tutorials}, vol. 20, no. 4, pp. 3098-3130, 2018.
		
		\bibitem{backhaul-capacity}	
		D. P. Bertsekas, R. G. Gallager, and P. Humblet, {``Data networks,"}	\textit{Prentice-Hall International New Jersey}, 1992, vol. 2.
		
		\bibitem{server-capacity-2}
		H. Alameddine, M. H. K. Tushar, and C. Assi, {``Scheduling of Low Latency Services in Softwarized Networks,"}  \textit{IEEE Transactions on Cloud Computing}, no. pp, vol. , 2019.
		
		\bibitem{auxilary}
		P. Luong, F. Gagnon, C. Despins, and L. Tran, {``Joint Virtual Computing and Radio Resource Allocation in Limited Fronthaul Green C-RANs,"}  \textit{IEEE Transactions on Wireless Communications}, vol. 17, no. 4, pp. 2602-2617, April 2018.
		
		\bibitem{cvx}
		M. Grant and S. Boyd, {``CVX: Matlab Software for Disciplined Convex	Programming, version 2.1,"} [Online]    {http://cvxr.com/cvx}, Mar. 2014.
		
		\bibitem{MM-approximation}
		{Y. Sun, P. Babu and D. P. Palomar, {``majorization-minimization Algorithms in Signal Processing, Communications, and Machine Learning,"} \textit{IEEE Transactions on Signal Processing}, vol. 65, no. 3, pp. 794-816, 1 Feb.1, 2017.}
		
		\bibitem{pathgain}
		S. Kazemi and M. Rasti, {``Joint power control and sub-channel allocation for co-channel OFDMA femtocells,"} \textit{in Proceeding of 2016 IEEE Symposium on Computers and Communication (ISCC), Messina}, 2016, pp. 1171-1176.	
	\end{thebibliography}
\end{document}